\documentclass{aa}  

\usepackage{graphicx}
\usepackage{rotating}
\usepackage{array}
\usepackage{subcaption}
\usepackage{caption}
\usepackage{soul}
\usepackage{txfonts}
\usepackage[dvipsnames]{xcolor}
\usepackage{natbib,twoopt}
\usepackage[breaklinks=true]{hyperref} 
\bibpunct{(}{)}{;}{a}{}{,}             
\makeatletter
  \newcommandtwoopt{\citetads}[3][][]{\href{http://adsabs.harvard.edu/abs/#3}%
    {\def\hyper@linkstart##1##2{}%
     \let\hyper@linkend\@empty\citep[#1][#2]{#3}}}
  \newcommandtwoopt{\citepads}[3][][]{\href{http://adsabs.harvard.edu/abs/#3}%
    {\def\hyper@linkstart##1##2{}%
     \let\hyper@linkend\@empty\citep[#1][#2]{#3}}}
  \newcommandtwoopt{\citeyearads}[3][][]%
    {\href{http://adsabs.harvard.edu/abs/#3}
    {\def\hyper@linkstart##1##2{}%
     \let\hyper@linkend\@empty\citeyear[#1][#2]{#3}}}
\makeatother
\begin{document}


   \title{Determining stellar accretion rates from $Pa_{\alpha}$ and $Br_{\beta}$ emission lines with JWST NIRSpec.}

   \subtitle{Accretion of pre-main-sequence stars in NGC 3603}

      \author{Ciarán Rogers
          \inst{1},
          Guido de Marchi
          \inst{2},
          Bernhard Brandl
          \inst{1, 3}}

   \institute{Leiden Observatory, Leiden University,
              PO Box 9513, 2300 RA Leiden, The Netherlands\\
              \email{rogers@strw.leidenuniv.nl}
         \and
             European Space Research and Technology Centre, Keplerlaan 1, 2200 AG Noordwijk, The Netherlands\\
             \email{gdemarchi@rssd.esa.int}
         \and
             Faculty of Aerospace Engineering, Delft University of Technology, Kluyverweg 1, 2629 HS Delft, The Netherlands}

   \date{Received December 18, 2023; accepted January 1, ????}
   \titlerunning{NIR line derived accretion luminosities of pre-main-sequence stars.}
    \authorrunning{Ciarán Rogers, Guido de Marchi, Bernhard Brandl}

   \date{Received January 19 2024; accepted March 11 2024}

 
  \abstract
   {In this Letter, we present the first systematic spectroscopic measurements of the near-infrared (NIR) hydrogen recombination lines Paschen alpha ($Pa_{\alpha}$ $\lambda = 1.875 \mu m$) and Brackett beta ($Br_{\beta}$ $\lambda = 2.626 \mu m$), produced by pre-main-sequence (PMS) stars. Such stars include T Tauri and Herbig AeBe stars, located in the massive Galactic star-forming region NGC 3603. We used measurements obtained from JWST NIRSpec, using multi-object spectroscopy (MOS) mode. Based on the existing empirical relations between $L_{acc}$ and $L_{Br_{\gamma}}$ from the literature, we used our new measurements to formulate, for the first time, an empirical relationship between the accretion luminosity, $L_{acc}$, of the stars and the line luminosities, $L_{line}$, of both $Pa_{\alpha}$ and $Br_{\beta}$. These relationships are: $\log_{10}( \frac{L_{acc}}{L_{\odot}}) = 1.42 (\pm 0.18) \times \log_{10}( \frac{L_{Pa_{\alpha}}}{L_{\odot}}) + 3.33 (\pm 0.42)$ and $\log_{10}( \frac{L_{acc}}{L_{\odot}}) = 1.47 (\pm 0.18) \times \log_{10}( \frac{L_{Br_{\beta}}}{L_{\odot}}) + 4.60 (\pm 0.57)$. These new relationships are key to establishing rough estimates of the accretion rates for large samples of PMS stars with JWST.}

   \keywords{Stars: formation - (Stars:) circumstellar matter - Stars: pre-main-sequence
               }

   \maketitle
\section{Introduction}
\label{sec:intro}
The accretion luminosity, $L_{acc}$, of a star is a direct measurement of the gravitational energy released by infalling matter as it is accreted. If the mass, $M_{*}$, and radius, $R_{*}$, are known, then the mass accretion rate, $\dot{M}_{*}$, can be determined from $L_{acc}$. The mass accretion rate is a fundamental characteristic of star formation, setting important constraints on accretion and stellar evolution models \cite{calvet1999evolution}. It is also a crucial aspect of disk evolution and star and planet formation \citep{muzerolle1998brgamma, basri1989accretion}.
The accretion luminosity of a star is primarily radiated at ultraviolet (UV) and blue and optical wavelengths, producing excess emission above the intrinsic photospheric level of the star \citep[e.g.][]{calvet1998structure}. It is possible to measure this accretion luminosity directly by observing the excess UV and optical continuum emission and comparing it to an appropriate stellar photosphere. This is commonly done using U and B band photometry with complementary spectroscopy,\citep[e.g.][]{venuti2014mapping, herczeg2008uv, gullbring1998disk}.\\ 
Another approach to measuring the accretion luminosity is to use hydrogen recombination lines. The accretion luminosity has been shown to scale linearly with various recombination line luminosities. In \cite{muzerolle1998brgamma}, they found that the NIR hydrogen recombination line $Br_{\gamma}$  ($\lambda = 2.166 \mu m$) exhibited a tight scaling relationship with the accretion luminosity of 19 pre-main-sequence (PMS) stars. Such close relationships have since been found for a wide range of lines from hydrogen, as well as other elements \citep[e.g.][]{alcala2017x}.\\
The motivation behind using NIR recombination lines is two-fold. Firstly, NIR lines suffer from $\sim 10$ times less extinction compared to UV and optical wavelengths. 
This results in a higher signal-to-noise ratio (S/N), making it possible to measure the accretion rates of more deeply embedded and/or distant objects. Secondly, because the extinction is lower, uncertainties that arise from extinction corrections are also smaller.\\
The majority of NIR stellar accretion studies have been conducted from ground-based observatories. From the ground, the strong recombination lines $Pa_{\alpha}$ and $Br_{\beta}$ are not visible due to telluric contamination. Using the JWST Near InfraRed Spectrograph (NIRSpec), we were able to obtain moderate-resolution ($R=2700$) spectra for $100$ stars, $32$ of which we  classified as PMS stars. The grating-and-filter combination of $G235H/F170LP$ was used. This gives us access to  recombination lines that are not observable from the ground: $Pa_{\alpha}$, a line that is typically $\sim 12$ times brighter than $Br_{\gamma}$, and $Br_{\beta}$, typically $\sim 1.6$ times brighter than $Br_{\gamma}$. These lines provide a new means to measure the accretion luminosity of PMS stars in the era of JWST, offering higher S/N, coupled with the exquisite sensitivity of NIRSpec, thus allowing for deeper and more precise star formation studies. 
The layout of this paper is as follows. In Section \ref{sec:targets}, we discuss the sample and our target selection criteria. In Section \ref{sec:prep}, we discuss the data processing steps, including the data reduction, extinction correction, flux calibration, photospheric and veiling corrections, and the measurement of the recombination lines. In Section \ref{sec:results} we present our results and in Section \ref{sec:concl} we summarise our findings and present our conclusions.

\section{Targets}
\label{sec:targets}
The spectra were obtained as part of a NIRSpec Guaranteed Time Observations (GTO) programme (ID=1225). The stars reside in the giant Galactic HII region NGC 3603, located $7 \pm 1$ kpc away \citep{melena2008massive, sung2004initial, pandey2000stellar}. The cluster's age has been estimated to be $1 - 3$ Myr \citep{pang2013origin, sung2004initial, melena2008massive}, with active ongoing star formation \cite{beccari2010progressive} (B10). The spectra were obtained using the MOS mode of NIRSpec, which utilises the Micro-Shutter Assembly (MSA). The MSA consists of $\sim 250\,000$ microscopic, operable doors that can be commanded open or closed, allowing for the simultaneous acquisition of dozens of spectra within a single exposure \cite{ferruit2022near}. These targets were chosen from an initial list of about $10\,000$ sources, originally observed with Hubble Space Telescope's (HST) Wide Field Camera 3 (WFC3), see (B10). Of the 100 sources observed with JWST, 60 were photometrically classified as PMS stars based on $H{\alpha}$ narrow band excess emission; the remaining 40 stars were classified as main-sequence (MS) stars.\\ 
With our new spectroscopic data, we have revised this classification. We have classified 32 of the 100 sources as PMS. This lower classification rate likely comes from our stricter criteria. Following our classification criteria, a source was deemed PMS based on the simultaneous presence of the recombination lines: $Pa_{\alpha}$, $Br_{\beta}$, and $Br_{\gamma}$, in emission above the chromospheric level, after a photospheric absorption correction had been performed (see Section \ref{subsubsec:veiling}). There are sources in our sample that exhibit only a subset of these lines in emission, in most cases, only $Pa_{\alpha}$. We have excluded these sources from our analysis simply because it is not possible to draw a relationship between $Br_{\gamma}$ and the other recombination lines if $Br_{\gamma}$ is not seen in emission. It is entirely possible that there are additional PMS sources in our sample, but within the scope of this study, we have focused on sources with strong recombination lines, unambiguously in emission. Of the 32 sources that we classified as PMS, 21 had also been classified in (B10) as PMS based on $H{\alpha}$ excess. The remaining 12 sources did not show evidence of $H{\alpha}$ excess at the time the observations were made. Differences in the nebular background subtraction, extinction correction and photospheric correction could explain the differences between our classification and (B10). There is also a $\sim 13$ year time gap between our NIRSpec measurements and the photometric measurements of (B10); thus, variability could also play a role. The sources are listed in Table \ref{tab:all_sources} with their positions, physical properties, recombination line equivalent widths (EWs), and accretion luminosities.\\
\begin{table*}
\small
    \centering
    \begin{tabular}{cccccccccc}
         \hline
         ID&  RA&  DEC&  $T_{eff}$ (K)&  Group &$M_*$ ($M_{\odot}$)&  EW $Pa_{\alpha}$ ($\AA$)&  EW $Br_{\gamma}$ ($\AA$)&  EW $Br_{\beta}$ ($\AA$)&  $\log_{10}(L_{acc})$ $(L_{\odot})$\\
         \hline
         \hline
         727& 168.8096466& -61.263581& 3800 $\pm 150$&  TT&0.5 $\pm 0.06$& -23.1 $\pm 2.2$& -0.8 $\pm 0.2$& -9.7 $\pm 1.9$& -1.42 $\pm 0.75$\\ 
         654&  168.7541941& -61.2514463&  4600 $\pm 150$&   TT&0.75 $\pm 0.07$&  -3.6 $\pm 0.4$&  -0.4 $\pm 0.1$&  -3.8 $\pm 0.7$&  -1.42 $\pm 0.72$\\ 
         4595&  168.8129136&  -61.2625111&  3600 $\pm 150$&   TT&0.47 $\pm 0.05$&  -22.2 $\pm 2$&  -2.4 $\pm 0.7$&  -26 $\pm 4.7$&  -1.4 $\pm 0.74$\\ 
         996&  168.7973597&  -61.2640824&  5200 $\pm 150$&   TT&0.97 $\pm 0.08$&  -8.9 $\pm 0.7$&  -0.5 $\pm 0.2$&  -2.9 $\pm 0.5$&  -1.11 $\pm 0.74$ \\ 
         1497&  168.7688519&  -61.25128&  6000 $\pm 100$&   TT&1.3 $\pm 0.04$&  -24.2 $\pm 2$&  -2.9 $\pm 0.5$&  -17.1 $\pm 2.9$&  -1.08 $\pm 0.71$ \\ 
         1457&  168.7310205&  -61.2672106&  5400 $\pm 100$&   TT&1.05 $\pm 0.04$&  -45.5 $\pm 3.8$&  -10.7 $\pm 1.3$&  -61 $\pm 10.5$&  -0.75 $\pm 0.69$ \\ 
         1038&  168.7957658&  -61.2643274&  5400 $\pm 150$&   TT&1.05 $\pm 0.08$&  -54.2 $\pm 4.6$&  -3.6 $\pm 0.5$&  -12.2 $\pm 2$&  -0.75 $\pm 0.68$ \\ 
         1339&  168.8197744&  -61.2630422&  4400 $\pm 150$&   TT&0.7 $\pm 0.06$&  -12 $\pm 1$&  -5.7 $\pm 0.7$&  -45 $\pm 7.8$&  -0.73 $\pm 0.69$ \\ 
         858&  168.7365507&  -61.2702015&  9600 $\pm 200$&   HAe/Be&3.3 $\pm 0.14$&  -82.6 $\pm 6.9$&  -7.6 $\pm 0.8$&  -33.5 $\pm 5.9$&  -0.67 $\pm 0.48$ \\ 
         1813&  168.8037358&  -61.246069&  5600 $\pm 150$&   TT&1.1 $\pm 0.08$&  -48 $\pm 4.1$&  -5.6 $\pm 0.7$&  -38 $\pm 6.7$&  -0.65 $\pm 0.69$ \\ 
         1442&  168.8120917&  -61.2671703&  4200 $\pm 150$&   TT&0.65 $\pm 0.06$&  -11 $\pm 0.9$&  -1 $\pm 0.2$&  -2.4 $\pm 0.4$& -0.65 $\pm 0.69$ \\ 
         1997&  168.8052566&  -61.2510052&  4400 $\pm 200$&   TT&0.7 $\pm 0.06$&  -14.8 $\pm 1.3$&  -1.3 $\pm 0.2$&  -3.2 $\pm 0.6$&  -0.61 $\pm 0.69$ \\ 
         1717&  168.8138352&  -61.2507679&  5800 $\pm 200$&   TT&1.2 $\pm 0.08$&  -24.6 $\pm 2.1$&  -3.4 $\pm 0.6$&  -6.7 $\pm 1.2$&  -0.61 $\pm 0.76$ \\ 
         4649&  168.7420813&  -61.2585345&  3600 $\pm 150$&   TT&0.47 $\pm 0.05$&  -128.7 $\pm 11.7$&  -10.6 $\pm 1.8$&  -59.3 $\pm 10.6$&  -0.53 $\pm 0.67$ \\ 
         1271&  168.7330151&  -61.2635352&  4600 $\pm 200$&   TT&0.75 $\pm 0.07$&  -70.6 $\pm 6$&  -9 $\pm 1.1$&  -17.5 $\pm 3.1$&  -0.47 $\pm 0.67$ \\ 
         2326&  168.7438995&  -61.2588048&  3800 $\pm 150$&   TT&0.5 $\pm 0.06$&  -114.3 $\pm 10.1$&  -25.7 $\pm 4.4$&  -86.3 $\pm 16.3$&  -0.41 $\pm 0.67$ \\ 
         1134&  168.7353906&  -61.2705277&  5600 $\pm 150$&   TT&1.13 $\pm 0.08$&  -125.1 $\pm 10.5$&  -10.1 $\pm 1.2$&  -45.4 $\pm 8.1$&  -0.37 $\pm 0.67$ \\ 
         2673&  168.7975961&  -61.2649162&  4400 $\pm 100$&   TT&0.7 $\pm 0.03$&  -34.2 $\pm 3$&  -3.5 $\pm 0.5$&  -7.3 $\pm 1.4$&  -0.36 $\pm 0.67$ \\ 
         1854&  168.7292443&  -61.2600537&  4200 $\pm 150$&   TT&0.75 $\pm 0.07$&  -225.1 $\pm 19.7$&  -22.8 $\pm 2.9$&  -127.9 $\pm 22.3$&  -0.36 $\pm 0.67$ \\ 
         2492&  168.7427501&  -61.2714228&  5000 $\pm 100$&   TT&0.9 $\pm 0.04$&  -245 $\pm 21.8$&  -57.6 $\pm 7$&  -79.2 $\pm 14.7$&  -0.27 $\pm 0.66$ \\ 
         977&  168.8092518&  -61.2623119&  5000 $\pm 150$&   TT&0.9 $\pm 0.07$&  -19.1 $\pm 1.6$&  -2 $\pm 0.3$&  -5.8 $\pm 1.1$&  -0.03 $\pm 0.67$ \\ 
         354&  168.7704651&  -61.2626355&  10600 $\pm 2000$&   HAe/Be&$4.02^{+1.7}_{-1.2}$&  -22.5 $\pm 2.1$&  -5 $\pm 1.3$&  -5.6 $\pm 1.3$&  0.07 $\pm 0.49$ \\ 
         853&  168.7643462&  -61.2488318&  5400 $\pm 100$&   TT&1.05 $\pm 0.04$&  -18.2 $\pm 1.5$&  -2.1 $\pm 0.3$&  -5.9 $\pm 0.9$&  0.15 $\pm 0.64$ \\ 
         2166&  168.7944201&  -61.2661562&  5000 $\pm 150$&   TT&0.9 $\pm 0.07$&  -69.1 $\pm 6.2$&  -6.5 $\pm 0.8$&  -15.3 $\pm 2.6$&  0.27 $\pm 0.64$ \\ 
         152&  168.7817328&  -61.2648809&  9400 $\pm 1800$&   HAe/Be&$4.33^{+1.8}_{-1.3}$&  -14.9 $\pm 1.2$&  -1.5 $\pm 0.2$&  -2.2 $\pm 0.4$&  0.42 $\pm 0.44$ \\ 
         1350&  168.8031179&  -61.2483281&  5000 $\pm 2000$&  TT/HAe &0.9 $\pm 0.7$&  -53.6 $\pm 4.4$&  -8.2 $\pm 1$&  -11.1 $\pm 2$&  0.66 $\pm 0.63$ \\ 
         1354&  168.7519521&  -61.2645637&  6600 $\pm 1400$&   TT/HAe&$2.18^{+ 0.90}_{-0.64}$&  -20.8 $\pm 1.7$&  -3.5 $\pm 0.4$&  -5 $\pm 0.8$&  0.9 $\pm 0.62$ \\ 
         823&  168.7623889&  -61.2710101&  7200 $\pm 1400$&   HAe&$3.11^{+ 1.3}_{-0.92}$&  -33.8 $\pm 2.9$&  -6 $\pm 0.7$&  -9 $\pm 1.5$&  1.79 $\pm 0.43$ \\ 
         238&  168.8151779&  -61.2550836&  7000 $\pm 1800$&   HAe/Be&$5.47^{+ 2.3}_{-1.6}$&  -76 $\pm 6.3$&  -13.5 $\pm 2.1$&  -23.6 $\pm 4.2$&  2.06 $\pm 0.39$ \\ 
         185&  168.7627924&  -61.2625057&  8200 $\pm 1800$&   HAe/Be&$5.06^{+ 2.1}_{-1.5}$&  -82.2 $\pm 6.8$&  -9.4 $\pm 1.1$&  -14.1 $\pm 2.3$&  2.13 $\pm 0.37$ \\ 
         469&  168.7941589&  -61.2779717&  8600 $\pm 1800$&   HAe/Be&$5.16^{+ 2.1}_{-1.5}$&  -216.9 $\pm 45.1$&  -16.2 $\pm 2$&  -17.2 $\pm 3.1$&  2.52 $\pm 0.36$ \\
         251& 168.7684956& -61.2531283& 6400 $\pm 1800$&  TT/HAe&$3.74^{+ 1.5}_{-1.1}$& -124.9 $\pm 10.6$& -13.7 $\pm 1.6$& -17.6 $\pm 3.1$& 2.75 $\pm 0.37$\\
    \end{tabular}
    \caption{\small{ Physical properties of the PMS stars considered in this analysis and the EWs of the three recombination lines. The sources have been ordered in terms of increasing accretion luminosity. The Group column indicates whether the sources are classified as T Tauri or Herbig AeBe. For the T Tauri stars, the temperatures, masses and associated uncertainties come directly from the Phoenix stellar models based on the five best fitting model spectra. For the Herbig AeBe stars, the temperatures and masses come from the photometry of (B10). The uncertainty in mass for each source has been calculated based on $M_{*} \divideontimes \sqrt{2}$. The uncertainty in temperature for the Herbig AeBe stars has been determined from the Phoenix models, such that the range of temperatures reproduces the upper and lower mass limit for each source.\\ 
    Note: Source 1350 could not be classified by spectroscopic or photometric means and so, a temperature of $5000$ K was assumed with an uncertainty of $2000$ K.}}
    \label{tab:all_sources}
\end{table*}
It is possible for recombination lines to be present in MS/evolved sources through chromospheric emission \citep{herbig1985chromospheric, strassmeier1990chromospheric, young1989study}. Chromospheric activity can be confused for weak accretion, as both mechanisms produce emission lines with small EWs. Accretion is usually associated with large EWs, for example: $H_{\alpha} \ge 10 \AA$. This corresponds to EWs for $Pa_{\alpha}$ and $Br_{\beta}$ of $1.2 \AA$ and $0.15 \AA,$ respectively, assuming case B recombination \cite{1987MNRAS.224..801H}. To avoid any misclassifications, we have taken a conservative approach and only considered sources with hydrogen recombination lines whose EWs are $\ge 3$ times what is expected from chromospheric activity.\\ 
In some of the PMS spectra, there are additional hydrogen recombination lines, from the Brackett series as well as the Pfund series. A full description of the PMS spectra will be given in (Rogers et al. in prep.).\footnote{Individual spectra can be made available by the authors upon request.}\\ 
We have spectrally classified the majority of the PMS stars in this study. A subset of sources could not be classified spectroscopically as they lack any absorption lines. These PMS stars have temperatures masses estimated photometrically from (B10), ranging from $2 \le \frac{M_{*}}{M_{\odot}} \le 5.5$, with an uncertainty of the order of $M_{*} \divideontimes \sqrt{2}$, placing them somewhere within the intermediate mass T Tauri, and Herbig AeBe star range. The large uncertainties on mass and temperature prevent us from firmly classifying these sources. In Table \ref{tab:all_sources}, we refer to these sources as either TT/HAe or HAe/Be stars.

\section{Data reduction}
\label{sec:prep}
\subsection{Spectral extraction with NIPS}
The data were reduced using the NIRSpec Instrument Pipeline Software (NIPS) \cite{NIPS}. NIPS is a framework for spectral extraction of NIRSpec data from the count-rate maps, performing all major reduction steps from dark current and bias subtraction to flat fielding, wavelength and flux calibration and spectral extraction, with the final product being the 1D extracted spectrum.
\subsection{Nebular background subtraction}
\label{subsec:subtr}
NGC 3603 is the optically brightest HII region in the Galaxy. As such, the nebular emission of NGC 3603 is extremely bright, in particular with respect to the recombination lines, which have typical EWs of $\sim 5000 \AA$. The nebular emission is also spatially variable across the area of a few micro-shutters (micro-shutter area = $0.46\arcsec \times 0.2\arcsec \sim 10^{-4} pc^2$). The removal of these nebular recombination lines from the stellar spectra was crucial in order to determine accurate accretion luminosities from the genuine stellar recombination lines.\\
Spectroscopy of the nebula was carried out simultaneously with the stellar spectroscopy. This was done by opening 'slitlets', consisting of three micro-shutters for each stellar source. The star was placed within the central micro-shutter, with the neighbouring shutters directly above and below it observing the adjacent nebular background. From the multiple nebular spectra per stellar source, we observed that the nebular emission lines fluxes typically differed by $ 7\%$ between adjacent micro-shutters. In 5 out of 32 PMS sources, the corresponding nebular emission line fluxes differed by $\ge 20\%$ between adjacent micro-shutters. In these cases, the measured nebular spectrum was likely not representative of the nebulosity within the central micro-shutter; hence, this could lead to over- or under-subtraction.\\
In order to account for the spatially variable nebulosity, a brightness scaling procedure was developed to subtract the nebular light from the stellar spectrum. In all of the nebular spectra, a bright helium doublet is present at $\lambda = 1.869 \mu m$. For our sample of stars, this line has a purely nebular origin, only appearing weakly in absorption (EW $\le 1 \AA$) in massive stars \cite{husser2013new}. The line flux of the helium doublet was measured in the unsubtracted stellar spectrum and the corresponding nebular spectrum was scaled and subtracted from the stellar spectrum, such that the helium line was fully removed. The typical scaling factor was $0.94 \pm 0.2$. The typical He I EW after subtraction was $ 0.011 \pm 0.16 \AA$.\\
We have estimated the uncertainty of this nebular subtraction by comparing the EW of the helium doublet to that of the hydrogen recombination lines of interest, namely $Pa_{\alpha}$, $Br_{\gamma}$, and $Br_{\beta}$. The helium and hydrogen lines exhibit a tight scaling relationship with an origin at $\sim 0\AA$. This indicates that by removing the helium line, the hydrogen lines have also been fully subtracted. The dispersion of these relationships for $Pa_{\alpha}$, $Br_{\gamma}$, and $Br_{\beta}$ were $8 \%$, $11 \%,$ and $17 \%,$ respectively. These dispersions were taken as the uncertainty of our subtraction for each line and were propagated through to the final line luminosities.
\subsection{Extinction correction}
\label{subsec:ext_corr}
The stellar spectra also needed to be corrected for interstellar extinction. The extinction characteristics of NGC 3603 have been explored in detail in (Rogers et al., submitted). This study utilised the hydrogen recombination lines of the nebular spectra. By studying the decrement of the Brackett recombination lines, it was possible to determine the NIR extinction law for each of our sources in NGC 3603, which allowed us to determine the reddening $E(B-V)$ and the absolute extinction $A(V)$ for each source. Using these values, the stellar spectra were corrected for extinction, and the uncertainties from $R(V)$ and $E(B-V)$ and $A(V)$ were propagated through to the uncertainty of the recombination line luminosities. The typical value of $R(V)$ for the region is $R(V) = 4.8 \pm 1.06$, corresponding to $E(B-V) = 0.64 \pm 0.27$ and $A(V) = 3.1 \pm 1.46$
\subsection{Flux calibration}
\label{subsec:flux_cal}
The NIRSpec calibration does not yet offer absolute flux calibrated spectra and so, this must be done with a photometric catalogue. We have used the non-contemporaneous $K_s$ band photometry of \cite{brandl1999low}, obtained with ISAAC at the VLT. To avoid the impact of temporal variability from the PMS sources, we used only the MS sources to determine the flux correction. By convolving our JWST MS stellar spectra with the ISAAC $K_s$ throughput filter, we found that our spectra were typically fainter than the corresponding ISAAC fluxes by a factor of $3.69 \pm 0.93$. We then applied this correction factor to the 32 PMS spectra. The uncertainty of this flux calibration was propagated through to the uncertainty of the line luminosities.

\subsection{Determining $L_{Pa_{\alpha}}$ and $L_{Br_{\beta}}$}
\label{subsec:measure_lines}
\subsubsection{Measuring the recombination line EWs}
\begin{figure}[h]
    \centering
    \includegraphics[width=1\linewidth]{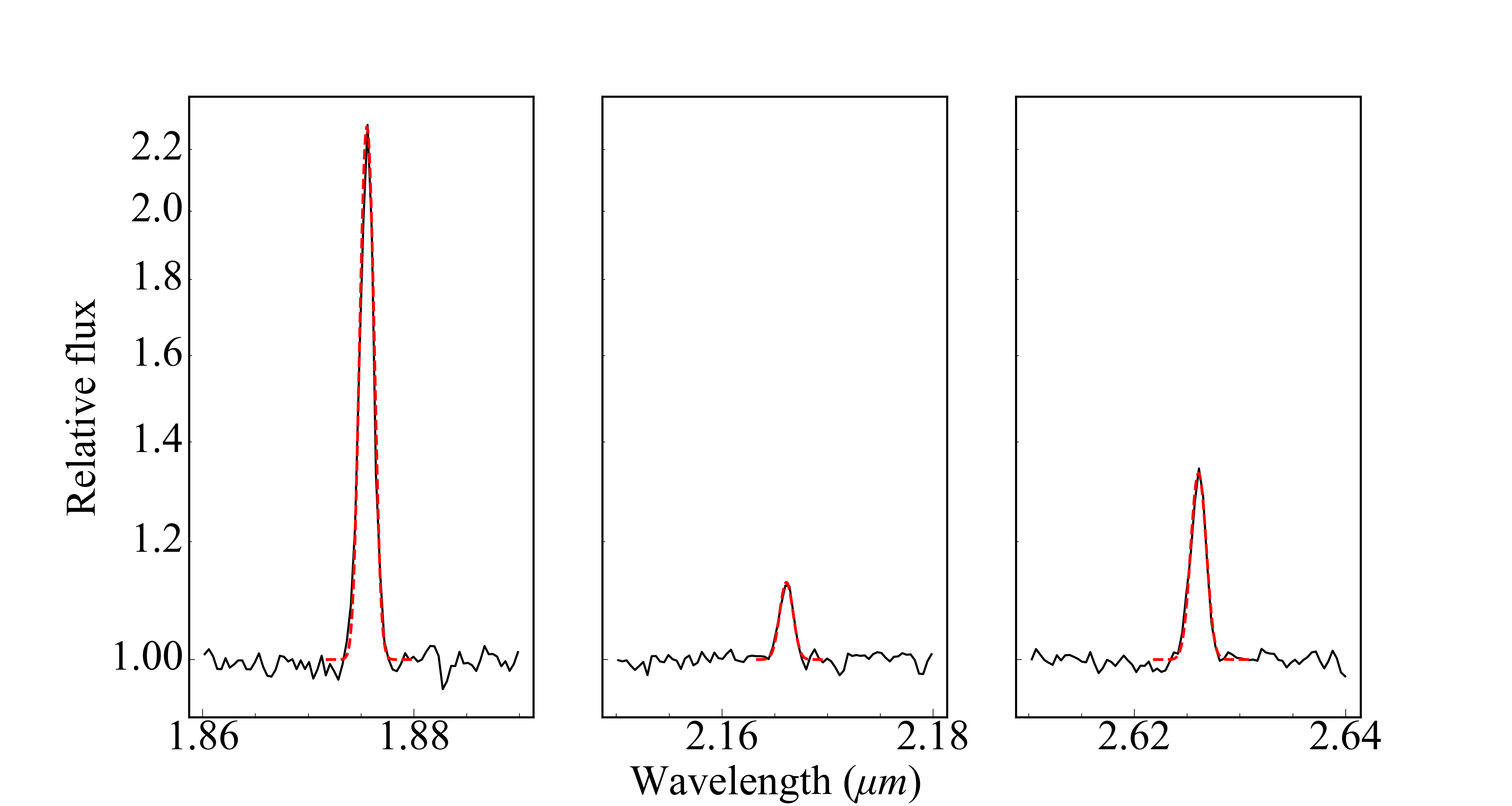}
    \caption{Three recombination lines: $Pa_{\alpha}$, $Br_{\gamma}$, and $Br_{\beta}$ are shown going from left to right. These lines come from the spectrum of source 977.}
    \label{fig:3_lines}
\end{figure}

A Monte Carlo approach was taken to measure the recombination lines, similarly to \citep{riedel2017young, ashraf2023h}. With this approach, each line was measured 1000 times by fitting a Gaussian to determine the the EW of the line. The EW was multiplied by the continuum to convert to a line flux. The line flux was allowed to vary with each iteration within its statistical and systematic uncertainties, under the assumption of independent and normally distributed uncertainties. The median of the 1000 fluxes was calculated and the uncertainty of the line flux was the standard deviation of the 1000 measurements. Figure \ref{fig:3_lines} shows the line profiles of $Pa_{\alpha}$, $Br_{\gamma}$, and $Br_{\beta}$ from the spectrum of source $977$, an intermediate S/N source in our sample. A full grid of line profiles is provided in Section \ref{app:emmis_line_profiles}.
\subsubsection{Correcting for photospheric absorption and veiling}
\label{subsubsec:veiling}
\begin{figure*}[h]
    \centering
    \includegraphics[width=1\linewidth]{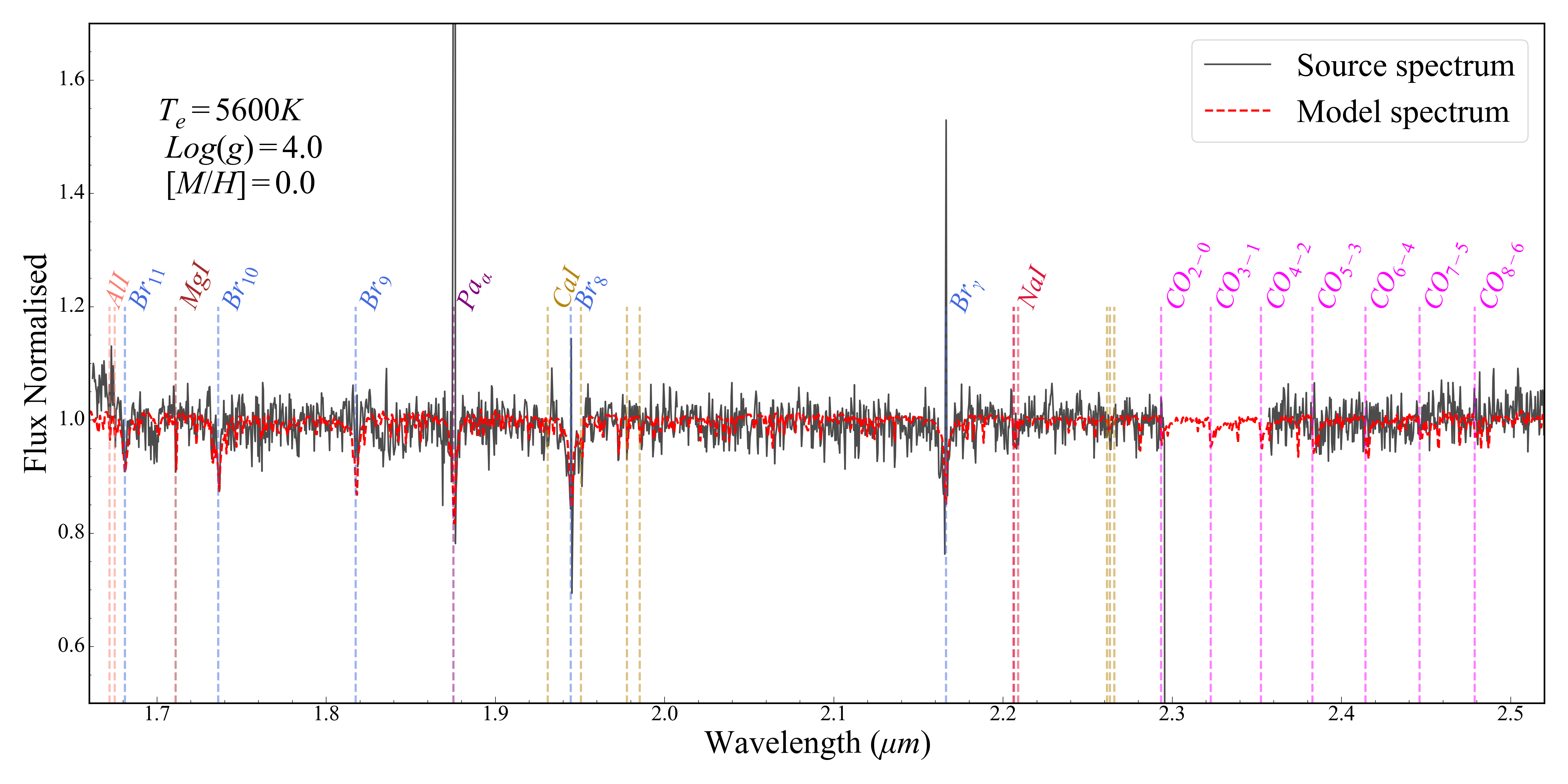}
    \caption{Best fitting Phoenix model spectrum. The source spectrum is shown as the black solid line. The best-fit model spectrum is shown as the red dashed line.}
    \label{fig:model_fit}
\end{figure*}

Before converting the EW to a flux, each EW needed to be corrected for photospheric absorption and veiling. To correct for photospheric absorption, the spectral type had to be determined. To do so, the Phoenix stellar models from \cite{husser2013new} were fit to the spectra using a minimum $\chi^2$ approach. A combination of hydrogen absorption lines, as well as metal absorption lines from $Na I$, $Ca I$, $Al I,$ and $Mg I$ were used to fit the spectra. An example fit is shown in Figure \ref{fig:model_fit}.
The uncertainties in $T_{eff}$, $Log(g),$ and metallicity $[M/H]$ were determined from the standard deviation of the five best fitting models. Once the spectral type was known, the EWs of $Pa_{\alpha}$, $Br_{\beta}$, and $Br_{\gamma}$ were measured in the model spectra. The EW uncertainties were also determined from the standard deviation of the five best fitting model spectra. For spectra where no absorption lines were detected, we used the spectral types determined by (B10), with their associated uncertainties.\\
The approach from \cite{lopez2006accretion} was used to simultaneously corrected the spectra for absorption and veiling,
\begin{equation} \label{veiling_form}
    \centering
    EW_{circ} = EW_{obs} - EW_{photo} \times 10^{-0.4 \times \Delta m_K}
\end{equation}
where $EW_{circ}$ is the true EW of the line emission from the circumstellar disk; $EW_{obs}$ is the measured EW of the line; $EW_{photo}$ is the EW of the photospheric absorption component; $\Delta m_K$ is the absolute value of the difference between the scaled Phoenix model, $K_s$, band magnitude and the observed $K_s$ band magnitude. To determine $\Delta m_K$, the extinction-corrected $K_s$ band magnitude of the observed spectrum was subtracted from the $K_s$ band magnitude of an appropriate Phoenix stellar model, scaled to the distance of NGC 3603. The absolute value of this difference was used for $\Delta m_K$. The corrected EWs were converted to fluxes as described above and, finally, to luminosities by scaling to the distance of $7 \pm 1$ kpc, and propagating the uncertainty of the distance.


\section{Results}
\label{sec:results}

\subsection{Determining the accretion luminosity relationship for $Pa_{\alpha}$ and $Br_{\beta}$}
\label{subsec:acc_lum}
\begin{figure*}[h!] 
\begin{subfigure}{0.5\linewidth}
    \centering
    \includegraphics[width=1\textwidth]{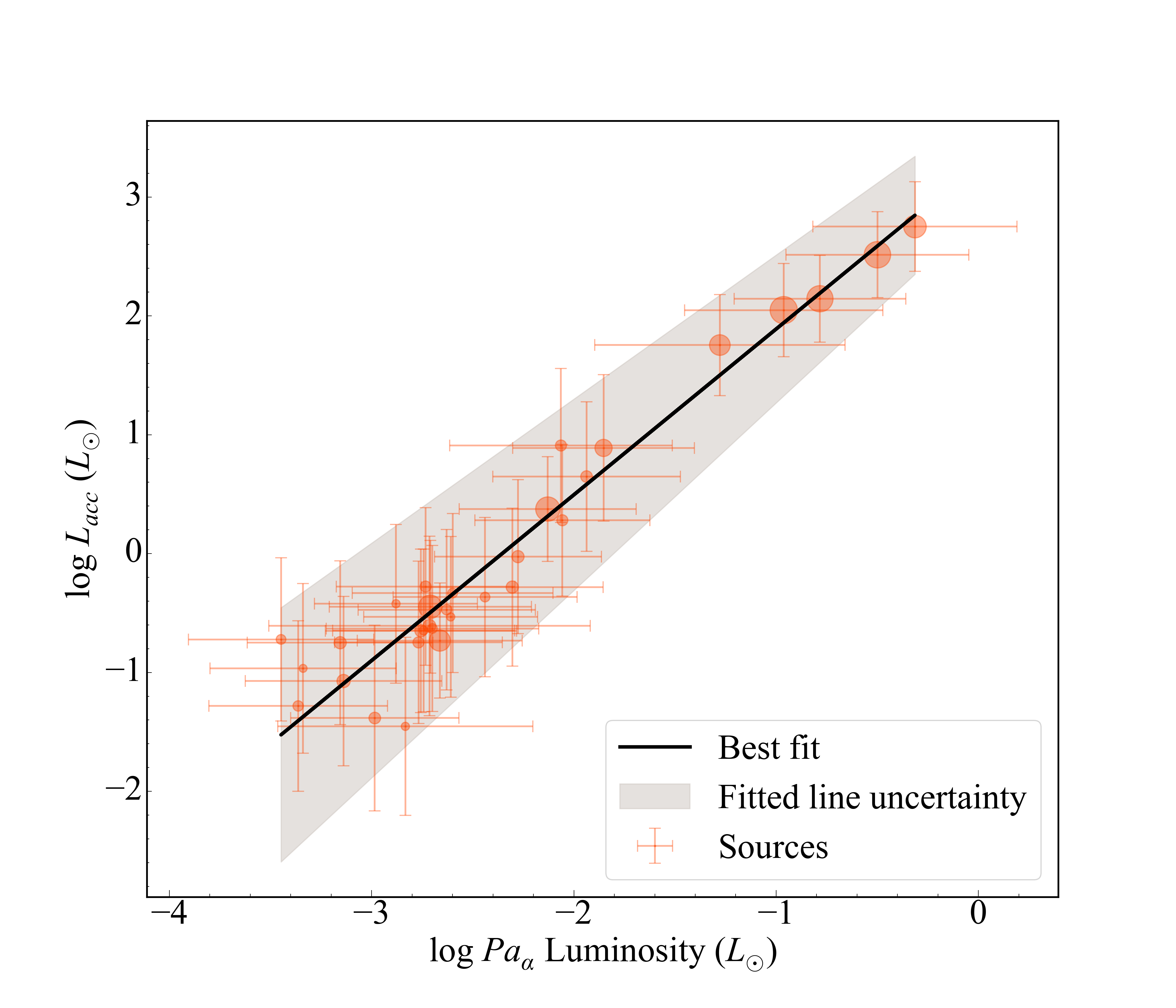}
\end{subfigure}
\quad
\begin{subfigure}{0.5\linewidth}
    \centering
    \includegraphics[width=1\textwidth]{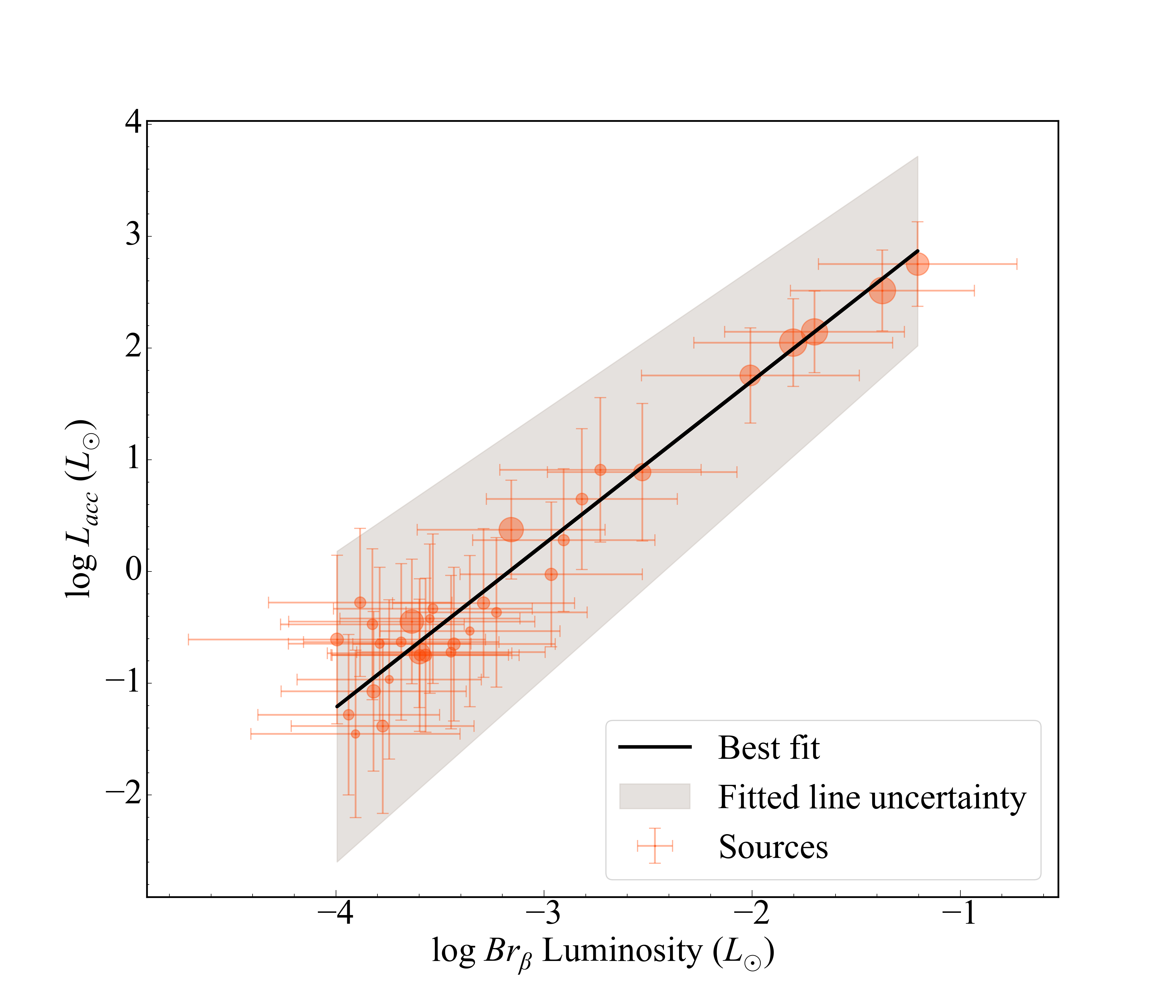}
\end{subfigure}
        
\caption{Relationship between $L_{acc}$ with $Pa_{\alpha}$ and $Br_{\beta}$. The size of the circle scales with the mass of the source. The black line is the line of best fit, determined via ODR. The grey shaded band is the uncertainty of the best fit line. Panel a: The scatter between data and the line of best fit is $\sigma = 0.27$. Panel b: The scatter between data and the line of best fit is $\sigma = 0.27$.}
\label{fig:pa_br_relation}
\end{figure*}

To determine the new empirical relationships between $L_{acc}$ and $Pa_{\alpha}$ and $Br_{\beta}$, we utilised two separate, existing relationships for $Br_{\gamma}$, determined by \cite{alcala2017x} for T Tauri stars and \cite{fairlamb2015spectroscopic} for the Herbig AeBe stars, although both relationships actually agree with each other within their uncertainties. These relationships are given in Equation \ref{form0} and \ref{form1}, respectively:
\begin{equation} \label{form0}
    \centering
    \log \; \frac{L_{acc}}{L_{\odot}} = 1.19 (\pm 0.1) \; \log \; \frac{L_{Br_{\gamma}}}{L_{\odot}} + 4.02 (\pm 0.51),
\end{equation}

\begin{equation} \label{form1}
    \centering
    \log \; \frac{L_{acc}}{L_{\odot}} = 1.30 (\pm 0.09) \; \log \; \frac{L_{Br_{\gamma}}}{L_{\odot}} + 4.46 (\pm 0.23).
\end{equation}
Using the relationships in Equations \ref{form0} and \ref{form1}, we calculated the accretion luminosities of our PMS stars based on their $Br_{\gamma}$ line luminosities.\\
In order to derive the relationships for $Pa_{\alpha}$ and $Br_{\beta}$, we fit lines between the accretion luminosities and the two recombination line luminosities. The best fitting lines are shown in Figure \ref{fig:pa_br_relation}. These fits provide the new empirically derived relationship between the accretion luminosity and $Pa_{\alpha}$ and $Br_{\beta}$. These relationships are given in Equations \ref{form2} and \ref{form3}:\\ 

\begin{equation} \label{form2}
    \centering
    \log_{10} \; (\frac{L_{acc}}{L_{\odot}}) = 1.42 (\pm 0.18) \; \log_{10} \; (\frac{L_{Pa_{\alpha}}}{L_{\odot}}) + 3.33 (\pm 0.42),
\end{equation}

\begin{equation} \label{form3}
    \centering
    \log_{10} \; (\frac{L_{acc}}{L_{\odot}}) = 1.47 (\pm 0.18) \; \log_{10} \; (\frac{L_{Br_{\beta}}}{L_{\odot}}) + 4.60 (\pm 0.57).
\end{equation}

The uncertainties of both the slope and y-intercept are dominated by systematic uncertainties. These systematics arise from the absolute flux calibration, the distance to NGC 3603, and the uncertainties inherited from Equations \ref{form0} and \ref{form1}. Taking Equation \ref{form2} for instance, the uncertainties can be broken down into their statistical and systematic components, with statistical uncertainty in square brackets and systematic uncertainty in angle brackets: $(\pm 0.18) \rightarrow [\pm 0.006] + \langle \pm 0.174 \rangle$ and $(\pm 0.42) \rightarrow [\pm 0.012] + \langle \pm 0.408 \rangle$.\\

\section{Conclusions}
\label{sec:concl}
We present the first empirical relationship between $L_{acc}$ and $L_{Pa_{\alpha}}$ and $L_{Br_{\beta}}$ by making use of the existing relationships between $L_{acc}$ and $L_{Br_{\gamma}}$ from \cite{alcala2017x} and \cite{fairlamb2015spectroscopic}. These new empirical relationships should serve star formation studies in the era of JWST, where simultaneous access to these strong NIR lines has been made available for the first time. This will allow for deeper and wider star formation studies at larger distances than has been previously possible.\\

\section{Acknowledgements}
We would like thank the referee for their careful consideration of this letter, and for their helpful comments which has improved its quality. We are also grateful to Juan Alcalá and Katia Biazzo for their invaluable help regarding the veiling and photospheric corrections for our data.


\bibliographystyle{aa} 
\bibliography{references.bib} 
\clearpage
\appendix
\section{Emission line profiles}
\label{app:emmis_line_profiles}
\begin{figure*}[hbt!]
    \centering
    \onecolumn 
    \includegraphics[width=1\linewidth]{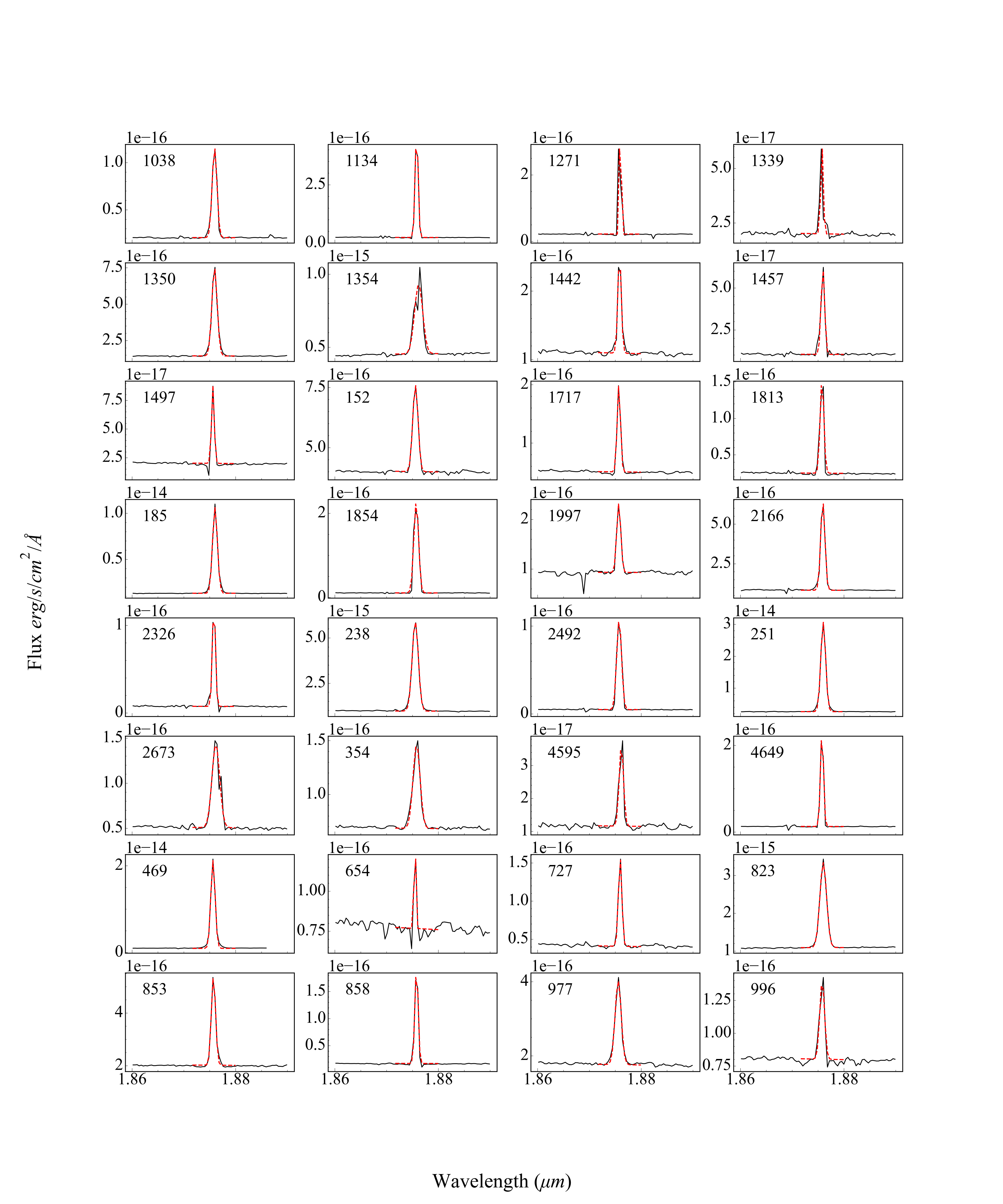}
    \caption{$Pa_{\alpha}$ emission line profile for each source. The best fitting Gaussian profile is shown as a red dashed line.}
\end{figure*}

\begin{figure*}[h]
    \centering
    \includegraphics[width=1\linewidth]{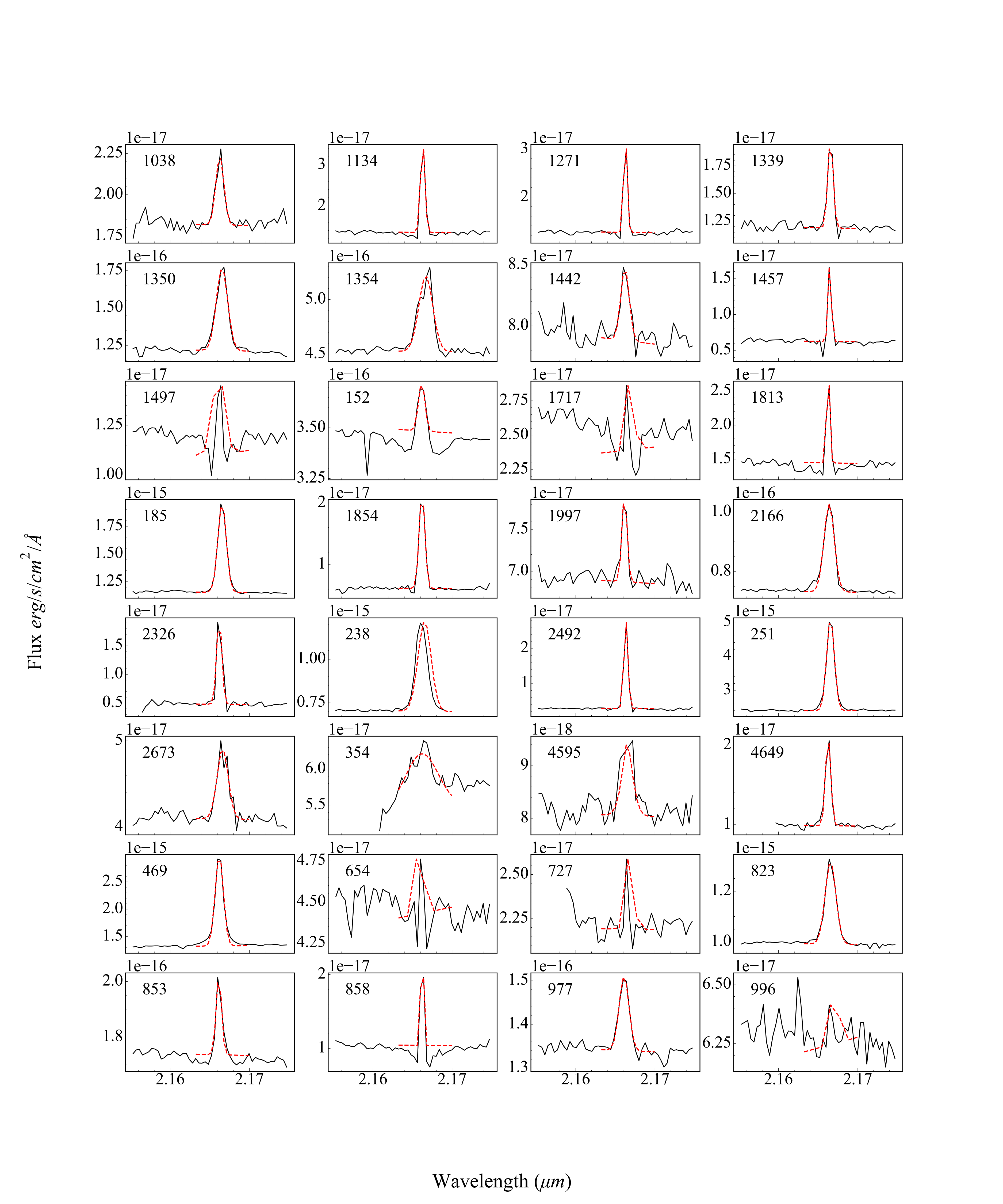}
    \caption{$Br_{\gamma}$ emission line profile for each source. The best fitting Gaussian profile is shown as a red dashed line.}
\end{figure*}

\begin{figure*}[h]
    \centering
    \includegraphics[width=1\linewidth]{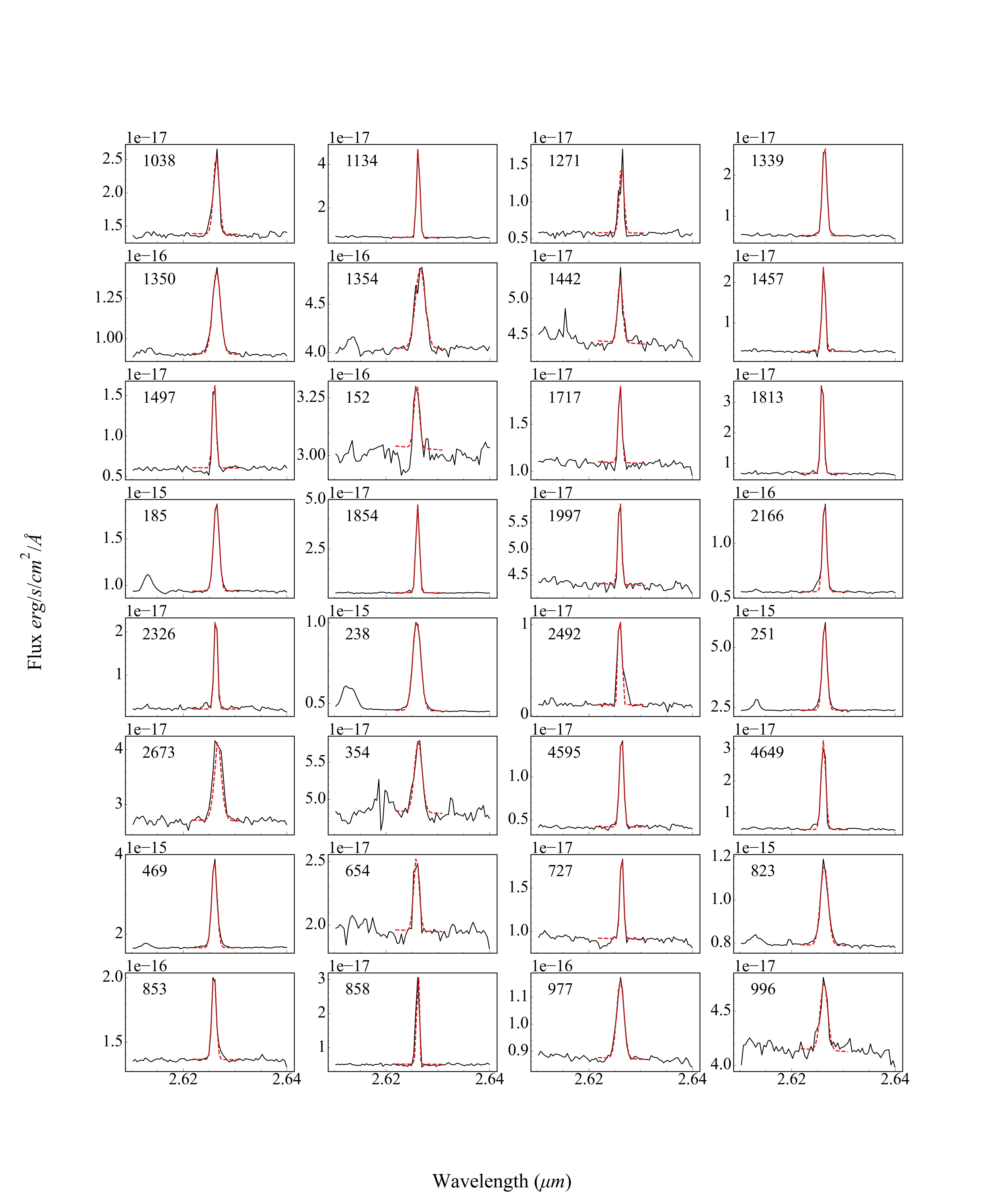}
    \caption{$Br_{\beta}$ emission line profile for each source. The best fitting Gaussian profile is shown as a red dashed line.}
    \label{fig:brb}
\end{figure*}

\end{document}